\documentclass[letterpaper]{aa}
\usepackage{epsf}
\usepackage{bm}
\usepackage{amsmath}
\usepackage{amssymb}
\usepackage{natbib}

\newcommand{\gcc}{\mbox{ g cm$^{-3}$}}
\newcommand{\beq}{\begin{equation}}
\newcommand{\eeq}{\end{equation}}
\newcommand{\bea}{\begin{eqnarray}}
\newcommand{\eea}{\end{eqnarray}}
\newcommand{\req}[1]{Eq.\ (\ref{#1})}

\newcommand{\Msun}{M_\odot}
\newcommand{\R}{\object{SAX J1808.4--3658}}

\begin{document}

\title{Thermal states of coldest and hottest neutron stars 
        in soft X-ray transients}
\titlerunning{Thermal states of neutron stars in SXTs}
\authorrunning{Yakovlev et al.}

\author{
  D. G. Yakovlev\inst{1} \and 
  K. P. Levenfish\inst{1,4} \and 
  A. Y. Potekhin\inst{1,4} \and
  O. Y. Gnedin\inst{2} \and
  G. Chabrier\inst{3}
}

\institute{Ioffe Physico-Technical Institute,
    Politekhnicheskaya 26, 194021 St.~Petersburg, Russia
    \and
    Space Telescope Science Institute,
     3700 San Martin Drive, Baltimore, MD 21218, USA
    \and
    Ecole Normale Sup\'erieure de Lyon
    (C.R.A.L., UMR CNRS No.\ 5574),
    46 all\'ee d'Italie, 69364 Lyon Cedex 07, France
    \and
    Isaac Newton Institute of Chile, St.~Petersburg Branch, Russia}

\offprints{A. Y. Potekhin,
    \email{palex@astro.ioffe.ru}}
    
\date{Received 13 August 2003 / Accepted 8 October 2003}

\abstract{
We calculate the thermal structure and quiescent
thermal luminosity of accreting neutron stars
(warmed by deep crustal heating in 
accreted matter) in soft X-ray transients (SXTs). We consider
neutron stars with nucleon and hyperon cores and
with accreted envelopes. It is assumed that an envelope
has an outer helium layer (of variable depth)
and deeper layers of heavier elements, either with iron
or with much heavier nuclei (of atomic weight $A\gtrsim100$) on the top
\citep{HZ90,HZ03}.
The relation between the
internal and surface stellar temperatures is obtained and fitted by
simple expressions. The quiescent
luminosity of the hottest (low-mass) and coldest (high-mass) neutron stars
is calculated, together with the ranges of its possible variations
due to variable thickness of
the helium layer. 
The results are
compared with observations of SXTs, particularly,
containing the coldest (\R) and the hottest (\object{Aql X-1}) neutron stars.
The observations of \R\ in a quiescent state on March 24, 2001 
\citep{campanaetal02}
can be explained only if
this SXT contains a massive neutron star
with a nucleon/hyperon core; a hyperon core
with a not too low
fraction of electrons is preferable.
Future observations may 
discriminate between the various 
models of hyperon/nucleon dense matter. The thermal emission
of \R\ is also sensitive to the models of plasma ionization in the
outermost surface layers and can serve 
for testing such models.
} 

\maketitle

\section{Introduction}
\label{sect:intro}

We study the thermal structure of accreting neutron stars in
soft X-ray transients (SXTs) --- close binaries with a low-mass
companion (e.g., \citealt{csl97}).  
Active states of SXTs are associated with
intense accretion energy release and accretion outbursts on
the neutron-star surfaces. These states are separated by long
periods of quiescence, when the accretion is
switched off or strongly suppressed.
As noticed by \citet{bbr98}, the spectrum of quiescent
emission is well fitted by a neutron-star
atmosphere model and may thus be of thermal origin,
being supported by the deep
crustal heating due to nuclear transformations
in the accreted matter.
 
Recently the thermal
structure and thermal emission of neutron stars in the SXTs
has been studied by \citet{bbc02},
taking into account that hydrogen burning in the surface
layers may proceed far beyond Fe, up to Te
(with nuclear mass numbers $A \sim 100$), via the rapid
proton capture process \citep{schatzetal01}. The
ashes of this burning have large nuclear charges $Z$, which
greatly reduces the thermal conductivity of accreted matter 
and increases the
internal stellar temperature  $T_\mathrm{in}$  for a given
effective surface temperature $T_\mathrm{eff}$
(or a given surface thermal luminosity $L_\gamma$). However, the heavy
nuclei may photodisintegrate in ``superbursts''
\citep{sbc03}, producing nuclei of the iron group, with
smaller $Z$. Moreover, recently \citet{woosleyetal03}
have performed new modeling of X-ray bursts with
updated physics input. Among many simulated X-ray
bursts, only one anomalous burst produced heavy nuclei
($A \sim 100$), while the other bursts produced nuclei with
$A \sim 60$. Taking into account a very wide range
of physical conditions in bursting neutron stars,
we consider both possibilities of burning to the
elements with $A \sim 60$ and $A \sim 100$.  

The nuclear ashes, left after bursts and superbursts
or after steady-state thermonuclear burning in the outermost layers, 
sink in the neutron star crust under the weight of newly
accreted matter. With increasing pressure, the sinking matter
undergoes a sequence of nuclear transformations
(particularly, pycnonuclear reactions), accompanied by
heat deposition (the so called {\it deep crustal heating}
in accreting neutron stars). \citet{HZ90} (hereafter HZ90)
studied these processes, starting from
the $^{56}$Fe ashes (see \citealt{bisnovatyi},
for references to some earlier work).  Recently the case of ashes
of much heavier elements has been considered 
by \citet{HZ03} (hereafter HZ03), with special attention
to $^{106}$Pd ashes. The
initial mass number $A$ strongly affects
the composition of accreted matter at
densities $\rho\lesssim10^{12}\gcc$,
less strongly  
at higher $\rho$, where pycnonuclear reactions operate,
and moderately affects
total crustal heat release
(1.45 MeV and 1.12 MeV per accreted nucleon for $A=56$
for 106, respectively).

Thermal states of neutron stars in SXTs
have been studied and compared with
observations by a number of authors (e.g.,
\citealp{ur01,colpietal01,rutledgeetal02a,bbc02,ylh}). 

Particularly, \citet{ylh} used a simplified model
of neutron-star thermal structure with the
iron heat-blanketing envelope, and employed
the relation between
$T_\mathrm{in}$ and $L_\gamma$
from \citet{PCY} (hereafter PCY). 
However, even a small mass $\Delta M$
of non-burned accreted 
H or He on the surface 
($\Delta M \gtrsim (10^{-18}-10^{-16})\,M_\odot$
depending on the surface temperature)  
noticeably increases the thermal conductivity
of the envelope (\citealt{CPY}; PCY). On the
contrary, the thermonuclear burning to high-$Z$ elements
(see above) decreases the thermal
conductivity. In the present paper, we take into account
both effects. First, we derive the relation between
$T_\mathrm{in}$ and $L_\gamma$ for an envelope composed of a
helium layer of arbitrary thickness and an underlying
heavy-element crust, described either by HZ90 or by HZ03
model. Second, we use this relation for calculating the thermal
states ($L_\gamma$
as a function of $\dot{M}$) of accreting neutron stars for
several neutron-star models with our fully
relativistic code of neutron-star thermal evolution. 
We consider five model
equations of state (EOSs) of matter in the neutron star cores
for two compositions of this matter --- nucleon matter
and nucleon-hyperon matter.
Finally, we compare the theoretical
results with observations of several SXTs in
quiescence. We make special emphasis on the hottest
and coldest neutron stars, in \object{Aql X-1} and \R, respectively.

\section{Thermal structure of heat-blanketing envelope}
\label{sect:th-str}

We calculate the thermal structure of a
transiently accreting neutron star in a quiescent state
using the same framework as for
an isolated neutron star with an accreted envelope
considered in PCY. A huge energy
released in the surface layers during the
active states of the system (heating of the surface by infall
of accreted matter and by thermonuclear burning of this matter)
is quickly carried to the surface by
the thermal conduction and radiated away
by the surface photon emission,
especially after a quiescence onset.
Thus, we solve numerically the thermal structure equation
in the stationary plane-parallel approximation
assuming no energy sources in the heat-blanketing
envelope.
In this way we calculate the temperature
distribution in the blanketing envelope,
with the thermal flux emergent
from warm stellar interiors. The equation to be solved is
\beq
  \frac{\mathrm{d}\,\log T}{\mathrm{d}\,\log P} = \frac{3}{16}\,
  \frac{PK}{g}\,\frac{T_\mathrm{eff}^4}{T^4},
\label{th-str}
\eeq
where $T$ is the temperature,
$P$ is the pressure,
$g$ is the surface gravity, and $K$ is the opacity.
The gravity is determined by the total gravitational
stellar mass $M$ and the circumferential stellar radius $R$:
$g=GM/(R^2 \sqrt{1-r_g/R})$, 
where $r_g=2GM/c^2=2.95\,M/M_\odot$ km
is the Schwarzschild radius.
The total opacity $K$ is
determined by the radiative (photon) opacity $K_\gamma$ and
the electron thermal opacity $K_e$:
$K^{-1} = K_\gamma^{-1} + K_e^{-1}$.
The effective temperature $T_\mathrm{eff}$
is defined by the Stefan law,
\begin{equation}
   L_\gamma = 4\pi R^2\sigma T_\mathrm{eff}^4,
\end{equation}
where $\sigma$ is the Stefan--Boltzmann constant, 
and $L_\gamma$
is the thermal surface luminosity in a local
neutron-star reference frame. The apparent luminosity
measured by a distant observer is
$L_\gamma^\infty=(1-r_g/R) \, L_\gamma$,
and the apparent surface temperature inferred
by the observer from the radiation spectrum is
$T_\mathrm{eff}^\infty=T_\mathrm{eff}\,\sqrt{1-r_g/R}$
(e.g., \citealt{Thorne}).

We adopt the standard 
outer boundary condition to Eq.~(\ref{th-str})
by equating $T_\mathrm{eff}$ to the temperature
$T_\mathrm{s}$ at the stellar ``radiative surface,''
which is found from the equation $P=2g/(3K)$.
\citet{spyz} have checked that the
replacement of this boundary condition by a more realistic condition,
which involves a solution to the radiative transfer problem in the
stellar atmosphere,
has almost no effect on the temperature distribution within
the blanketing envelope. We assume that the blanketing envelope
extends to the density $\rho_\mathrm{b}=10^{11}$ g cm$^{-3}$
and we integrate Eq.~(\ref{th-str}) from the surface 
to $\rho_\mathrm{b}$.
Thus, we define the internal neutron-star temperature as
$T_\mathrm{in}=T(\rho_\mathrm{b})$.

We assume further that the outermost neutron-star layer 
is composed of
$^4$He, which may be left as a non-consumed fuel, or  
may accrete on
the stellar surface after the last burst.
Actually, the outermost layer may be partly composed of hydrogen:
the replacement of He with H has very little effect 
on the thermal structure of the envelope (PCY).
We measure the thickness of this light-element layer by the parameter
\beq
   \eta = g_{14}^2\,\Delta M/M,
\eeq
where $\Delta M$ is the total mass of He (and H), and 
$g_{14}\equiv g/(10^{14}$ cm$^2$ s$^{-1}$) ($g_{14}$
ranges from $\sim 1$ to $\sim 4$, 
for different neutron-star masses and EOSs;
$g_{14}=2.43$ for a
``canonical'' neutron star of $M=1.4\,M_\odot$ and $R=10$ km).
The parameter $\eta$ is directly related to the pressure at the bottom
of the light-element envelope (see, e.g., \citealt{elounda}, Eq.\,(8)):
$P_\mathrm{He,max}=1.193\times10^{34}\,\eta$ dyn cm$^{-2}$.
\citet{bbc02} parameterized the thickness of the helium layer using
the column density $y$; it is related to $\eta$ as
\beq
   y=1.583\times10^{20}\,\frac{M}{M_\odot}\,\frac{\eta}{g_{14}^2 R_6^2}
   \mbox{~~g~cm}^{-2},
\eeq
where $R_6=R/10^6$ cm.
At $\eta\approx 4\times10^{-8}$, the helium layer reaches the density
$\rho\approx 10^9\gcc$, where it efficiently transforms into 
heavier elements
via pycnonuclear reactions and $\beta$-captures 
(e.g., \citealt{ml94}).
We do not consider higher values of $\eta$.

Under the helium layer, the crust is composed of
the burst ashes transformed under the action of beta captures,
emission and absorption of neutrons and (at $\rho\gtrsim10^{12}\gcc$)
pycnonuclear reactions.
We consider two
model compositions of these layers of the crust: the traditional
composition originating from the Fe
nuclei ($A=56$, HZ90) 
and the one originating from the Pd nuclei
($A=106$, HZ03).

As in PCY, we employ the OPAL radiative
opacity tables\footnote{%
http://www-phys.llnl.gov/V\_Div/OPAL/opal.html}
\citep{OPAL},
interpolated or extrapolated whenever necessary.
For the electron opacity
(inversely proportional to the electron thermal conductivity), we use
our code\footnote{%
http://www.ioffe.ru/astro/conduct/}
based on the theory presented by \citet{pbhy}.
In order to determine the density at a given pressure,
we use the Saumon--Chabrier EOS for the outer helium
layer \citep{scvh}, and the EOS of ideal relativistic electron plasma 
in the deep layers of the envelope
(using the fitting formulae of \citealt{CP}).

Note that our treatment of the thermal conductivity
of He and Fe matter takes into account the effects
of partial ionization, while the thermal
conductivity of Pd matter (HZ03) is calculated assuming full
ionization (because of the absence of appropriate
ionization models). Our estimates show
that the partial recombination of heavy ion plasma ($Z\gtrsim30$)
occurs (very roughly) at $\rho \lesssim 10^5$ g cm$^{-3}$
and $T \lesssim 10^7$ K (see Fig.~1 of PCY). 
If the main temperature
gradient in the blanketing envelope takes place
within the indicated domain, our results for the Pd/He 
envelope become
inaccurate. We expect, however, that the thermal
structure of such envelopes will not be too
different from the structure of equivalent He/Fe
envelopes (calculated with account for the partial ionization).
Nevertheless, strictly speaking, our calculations fail for cold neutron
stars ($T_\mathrm{eff} \lesssim 1.5 \times 10^6$ K) with the HZ03
envelopes and small amount of He on the surface
($\eta \lesssim 10^{-14}$). We have checked
this assumption by constructing artificial models of He/Fe envelopes
with fully ionized Fe. In a cold star with $T_\mathrm{in} \sim
6 \times 10^6$ K and $\Delta M/M \sim 10^{-20}$ such a
model gives the thermal surface luminosity about
2--3 times smaller than the accurate calculation.
We will see (Sect.\ \ref{results}) 
that the effects of partial ionization are
important for interpretation of observations.

\begin{figure}
\centering
\epsfxsize=86mm
\epsffile[80 220 520 640]{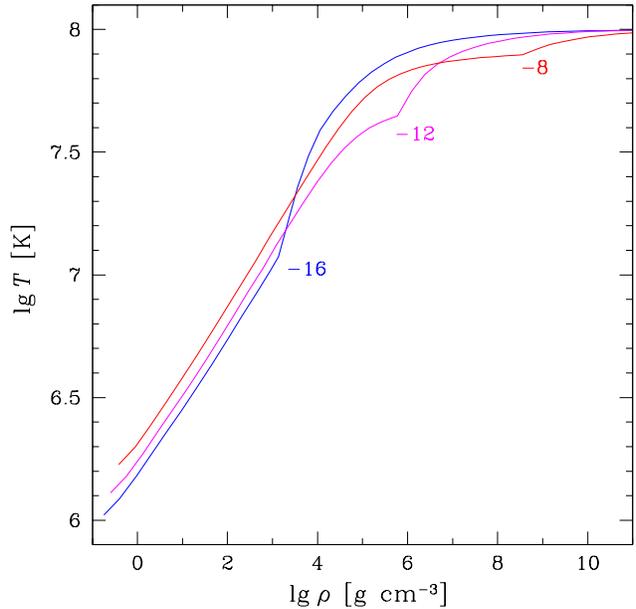}
\caption{
Temperature profiles (temperature versus density)
through an accreted heat-blanketing envelope
of the canonical neutron star ($M=1.4\, \Msun$, $R=10$ km).
The outer part of each model envelope (from the radiative surface 
to a cusp of a curve) is composed of $^4$He. The thickness
of the helium layer
is measured by parameter $\eta=g_{14}^2 \,\Delta M/M$: $\eta=10^{-16}$,
$10^{-12}$, and $10^{-8}$ (marked near the curves).
The composition of the inner part of the envelope (after the cusp)
is taken from \citet{HZ90}.
\label{fig:puhz}}
\end{figure}

Taking each model of the heavy-element accreted envelope (HZ90 and HZ03),
we have calculated $T_\mathrm{in}$ for a representative sample of 
$\sim 100$ pairs of
$T_\mathrm{eff}$ and $\eta$ values, in order to find the
$T_\mathrm{eff}$--$T_\mathrm{in}$ relation. The calculated
temperature profiles are exemplified in Fig.~\ref{fig:puhz}
(for $\eta=10^{-16}$,
$10^{-12}$, and $10^{-8}$). As
explained in PCY, the light-element layer of the envelope (He in our
case) provides a lower thermal insulation than the heavy-element one.
This causes a sharper
temperature drop from the inner isothermal layers to the
light/heavy-element interface. The interface produces a cusp on
each profile in Fig.~\ref{fig:puhz}. Outside the interface, the
temperature gradient is smaller than just inside. Therefore, the
thicker the He layer, the higher the effective temperature and
the emitted radiation flux, for a given $T_\mathrm{in}$.

\begin{figure}\centering
\epsfxsize=86mm
\epsffile[80 220 520 640]{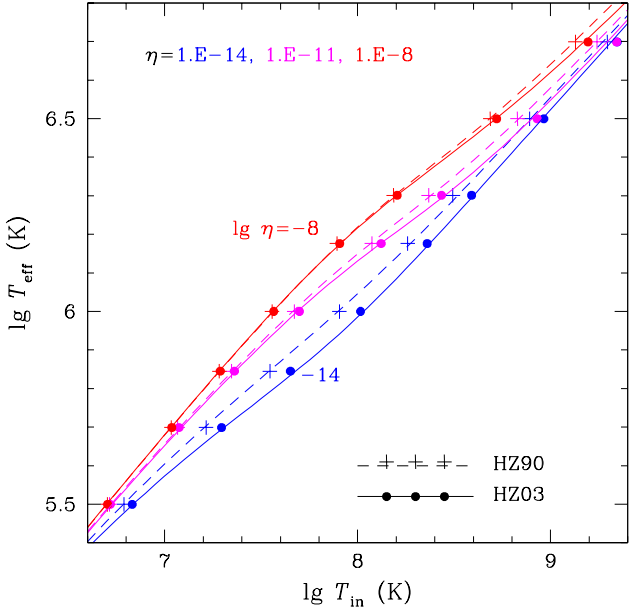}
\caption{
Effective temperature of
the canonical neutron star as a function of the internal
temperature
for the HZ90 composition of the envelope
(dashed lines) and for
the HZ03 composition (solid lines).
\label{fig:fit_hz}}
\end{figure}

As expected, the $T_\mathrm{eff}$--$T_\mathrm{in}$ relation in
the HZ90 case is almost the same as for a light-element
accreted envelope on top of the Fe layer considered by PCY (with
small corrections derived by \citealt{pycg}).
The replacement of H by He has little effect
on the $T_\mathrm{eff}$--$T_\mathrm{in}$ relation;
the effect is completely negligible
if $T_\mathrm{in}$ exceeds the temperature of hydrogen thermonuclear
burning ($\sim 4\times10^7$ K, \citealt{Ergma}).
However, the replacement of the HZ90 crust by the HZ03 crust
noticeably affects the 
$T_\mathrm{eff}-T_\mathrm{in}$ relation, as illustrated in
Fig.~\ref{fig:fit_hz}. In this figure, the solid lines and filled
circles show the dependence of $T_\mathrm{eff}$ on $T_\mathrm{in}$ for
the HZ03 accreted crust,
while the crosses and dashed
lines show this dependence for the HZ90 crust. 
The symbols demonstrate the results of numerical calculations,
while the lines correspond to the following fitting expressions:
\begin{subequations}
\label{fit}
\bea
   T_\mathrm{eff,6} &=& \left[ \gamma\,T_\mathrm{full,6}^4 +
   (1-\gamma)\,T_\mathrm{non,6}^4 \right]^{1/4},
\\
      T_\mathrm{non,6} &=& g_{14}^{1/4}\,a_1\,T_\mathrm{in,9}^{a_2},
\\
      T_\mathrm{full,6} &=& 6.9\,g_{14}^{1/4}\,
      C^{1/4}\,T_\mathrm{in,9}^{0.62},
   \quad
      C =
      \frac{1+a_3\,T_\mathrm{in,9}^{a_4}}{1+a_5\,T_\mathrm{in,9}^{1.6}},
\\
    \gamma &=& (1+120\,T_\mathrm{in,9})^p,
\quad
     p=-a_6/\eta^{a_7}.
\eea
\end{subequations}
Here, $T_\mathrm{eff,6}=T_\mathrm{eff}/10^6$ K,
$T_\mathrm{in,9}=T_\mathrm{in}/10^9$ K,
and the parameters $a_i$ ($i=1,2,\ldots,7$) are given in
Table~\ref{tab:param};
$T_\mathrm{non,6}$ and $T_\mathrm{full,6}$ refer to
$T_\mathrm{eff,6}$ in the absence
of a helium layer and 
in the presence of most massive helium layer,
respectively.
If $5\times10^6$~K $< T_\mathrm{in} < 10^9$ K, the difference between
the calculated and fitted values of $T_\mathrm{eff}$ does not exceed 4\%
for $\eta$ values used in the calculations: from 0 to $10^{-6}$ for the
HZ90 composition and from $10^{-16}$ to $10^{-6}$ for the HZ03
composition
(although 
the interval of $\eta$, where our physical model is applicable,
is smaller,
as explained above).

\begin{table}[!t]
\centering
\caption[]{Parameters of the fit (\protect\ref{fit})}
\label{tab:param}
\begin{tabular}{lll}
\hline\hline
\noalign{\smallskip}
& HZ90 & HZ03 \\
\hline
\noalign{\smallskip}
$a_1$ & 2.87 & 2.67 \\
$a_2$ & 0.534 & 0.560 \\
$a_3$ & 2.203 & 2.596 \\
$a_4$ & 1.349 & 1.136 \\
$a_5$ & 8.328 & 9.838 \\
$a_6$ & 0.01609 & 0.00967 \\
$a_7$ & 0.1378 & 0.1590 \\
\noalign{\smallskip}
\hline\hline
\end{tabular}
\end{table}

\section{Thermal states of transiently accreting neutron stars}
\label{sect:th-states}

\subsection{Observations}
\label{observations}

We will calculate the quiescent thermal luminosity
$L_\gamma^\infty$ of neutron stars in SXTs
versus the mean mass accretion rate $\dot{M}$ 
(from $10^{-15}$ to $10^{-9}$  $\Msun$ yr$^{-1}$) and compare 
the results with observations. Let us outline the observations first.
 
We take the same observational basis as in \citet{ylh}.
It consists of five SXTs
whose quiescent thermal luminosity and mean mass accretion
rate are measured or constrained (Fig.~\ref{fig3}):
\object{Aql X-1} \citep{rutledgeetal00,rutledgeetal02b}, 
\object{Cen X-4} \citep{csl97,rutledgeetal01}, 
\object{4U 1608--52} \citep{csl97,rutledgeetal99}, 
\object{KS 1731--26} \citep{wijnandsetal02,rutledgeetal02a}, 
and \R\ \citep{bc01,campanaetal02}. We adopt
the same data 
(see \citealt{ylh} for details) but 
analyze  more carefully the
hottest and coldest sources, 
\object{Aql X-1} and \R.  The data are rather
uncertain and are plotted thus as thick crosses.
The estimated values of $\dot{M}$ for Cen X-1 and KS 1731--26
are likely to be the upper limits,
as indicated by arrows.

Note that the quiescent luminosity 
of SXTs may vary from one observation to another. It is not clear
if this variability is associated with the thermal
radiation component or the non-thermal (power-law)
one, not related to the radiation emergent from neutron star interiors. 
For instance, \citet{rutledgeetal02b} report temporal variability
of radiation from \object{Aql X-1} over five months in quiescence
after an outburst in November 2000.
They fitted the thermal component of 
X-ray spectra with the hydrogen atmosphere
model assuming a constant neutron-star radius. 
In the beginning of the quiescent
period the thermal luminosity was estimated
as $L_\gamma^\infty \approx 9.3 \times 10^{33}$ erg s$^{-1}$.
It decreased to about $5.3 \times 10^{33}$
erg s$^{-1}$ in three months,  
increased to
about $6.3 \times 10^{33}$ erg s$^{-1}$ in the next month,
and stayed constant in the last month. 
In Fig.\  \ref{fig3} 
the observational error box is centered at the minimum value,
$L_\gamma^\infty \approx 5.3 \times 10^{33}$ erg s$^{-1}$,
while the upper value of the errorbar approximately
corresponds to the maximum value of $L_\gamma^\infty$.
Recently, \citet{cs03} have proposed another interpretation
of the same data. They assume that the quiescent thermal luminosity
stays constant while the temporal variability of the \object{Aql X-1}
radiation is associated with the variable nonthermal radiation
component (which arises outside the neutron star, for instance,
due to interaction of a pulsar wind 
with an accretion flow from the companion star).
Their value of the quiescent thermal luminosity
(redshifted for a distant observer) is in good agreement
with the central value of the observational errorbar in Fig.\ \ref{fig3}.   

We will also be interested in \R.
\citet{campanaetal02} report its  
\textit{XMM-Newton} observation
in a quiescent state on March 24, 2001, when
it was extremely weak. The spectrum of its 
emission was well fitted by power-law. It is most likely that the radiation
originates in a neutron-star magnetosphere or in a surrounding accretion
disk and is not related to the surface thermal emission. Thus, 
\citet{campanaetal02} detected 
no thermal
emission but 
obtained an upper limit on
$L_\gamma^\infty$ in that quiescent state.
The limit depends on the model of the thermal radiation
(black-body or hydrogen atmosphere model) and on
assumed neutron-star parameters (mass, radius, distance) and
varies from $\sim 7.5 \times 10^{29}$ erg s$^{-1}$ to
$2.5 \times 10^{31}$ erg s$^{-1}$. 

We consider two possible 
values (Figs.\ \ref{fig3} and \ref{fig5}) of the upper limit on $L_\gamma^\infty$
discussed by \citet{campanaetal02}. 
They seem to be more realistic than several other possible
upper limits mentioned by these authors.
The first value, $L_\gamma^\infty
=1.5 \times 10^{30}$ erg s$^{-1}$, is the upper
limit on the unabsorbed bolometric thermal luminosity
with the black-body (BB) spectrum. 
It is obtained from
the 90\% upper limit on the unabsorbed bolometric black-body flux
of $\sim 2 \times 10^{-15}$ erg s$^{-1}$ cm$^{-2}$. 
To translate the flux
to the luminosity we adopt the
distance to the source, 2.5 kpc, cited by these authors.
The second value, $L_\gamma^\infty
=7.5 \times 10^{29}$ erg s$^{-1}$, is the upper
limit on the unabsorbed bolometric thermal luminosity
obtained with the neutron-star hydrogen atmosphere (NSA) model. 

As mentioned, e.g., by \citet{campanaetal02} and \citet{ylh}, 
the quiescent thermal luminosity of \R\ is so low that 
the source should contain a very cold neutron star (with strong
neutrino emission). Particularly, as shown by \citet{ylh}, 
the data are compatible only with the model of a massive
neutron star whose core contains nucleons (and possibly hyperons)
and does not contain any exotic matter (pion- or kaon condensates
or quark matter, which would imply lower neutrino emission and
a noticeably hotter star). This restricts the class of possible EOSs
of dense matter to the models of nucleon/hyperon matter.
We extend the analysis of
\citet{ylh} by simulating thermal states of neutron stars
using an exact cooling code and several models
of nucleon/hyperon EOSs.

\subsection{Numerical simulations}
\label{code}

Following \citet{bbr98}, we assume that neutron stars in SXTs
are warmed up by deep crustal heating in accreted matter.
It is supposed that the heat released at their surface in active states
is radiated away, especially during quiescence,
and does not warm the stellar interiors.
Neutron stars in SXTs are thermally inertial objects, with typical relaxation
times $\sim 1-10$ kyr \citep{colpietal01}, while the mass accretion
rate $\dot{M}$ in SXTs varies on much shorter time scales.
Therefore, we will study
a global thermal state of a transiently accreting
neutron star by replacing a variable deep crustal
heating with the time-averaged heating determined by
the time-averaged accretion rate 
$\dot{M} \equiv \langle \dot{M} \rangle$. The deep-heating
power is
\begin{eqnarray}&&\!\!\!\!
    L_\mathrm{dh} = {Q \dot{M} \over m_\mathrm{N}}
 \nonumber\\&&\!\!\!
    \approx 6.03 \times 10^{33}\,
    \left({ \dot{M} \over 10^{-10}\; \Msun \; \mathrm{yr}^{-1}}\right) \,
    \left({ Q \over \mathrm{MeV}} \right)\;\; \mathrm{erg\;s}^{-1},
\label{Ldh}
\end{eqnarray}
where $m_\mathrm{N}$ is the nucleon mass, and
$Q$ is the total amount of heat released per one accreted nucleon
($Q=1.45$ MeV and 1.12 MeV for the HZ90 and HZ03 
models, respectively).

We simulate thermal states of accreting neutron stars
with our fully relativistic code of neutron-star thermal evolution
\citep{gyp01} by solving the
stationary thermal-balance equation:
\begin{equation}
     L_\mathrm{dh}^\infty(\dot{M})=L_\nu^\infty(T_\mathrm{in})
     + L_\gamma^\infty(T_\mathrm{eff}),
\label{thermal-balance}
\end{equation}
where $L_\nu^\infty$ is the neutrino luminosity of the star
as detected by a distant observer, and $L_\mathrm{dh}^\infty$
is the deep heating power for the same observer
(see, e.g., \citealt{ylh} for details). Since the mass
accretion rates in SXTs are determined with large uncertainties,
we do not make a difference between
$L_\mathrm{dh}$ and $L_\mathrm{dh}^\infty$ in \req{thermal-balance}.
The neutron star interior ($\rho > \rho_\mathrm{b}$)
is assumed isothermal (with the proper account of
the general relativistic effects). In this case
$L_\nu^\infty$ is a function of $T_\mathrm{in}$,
while $T_\mathrm{in}$ is related to $T_\mathrm{eff}$
(Sect.\ \ref{sect:th-str}).

We have updated our code in three respects. First,
we have incorporated the internal energy sources
associated with deep crustal heating. Second, we have
modified the relation between the surface and internal stellar
temperatures in accordance with the results
of Sect.\ \ref{sect:th-str}. Third, we have included new microphysics 
(neutrino emissivities and heat capacities) which allows us
to consider neutron-star cores containing $\Lambda$, $\Sigma^0$,
$\Sigma^-$, $\Sigma^+$, $\Xi^-$, and $\Xi^0$ hyperons
(in addition to neutrons, protons, electrons, and muons). 

The code calculates {\it heating curves},
the redshifted surface thermal luminosity of the star $L_\gamma^\infty$
(or the effective surface temperature $T_\mathrm{eff}^\infty$)
versus the mean mass accretion rate $\dot{M}$, to be
compared with observations. 
Examples are shown in Figs.\ \ref{fig3}--\ref{fig5}.

\subsection{Model equations of state of dense matter}
\label{EOS}

We will use five model EOSs in neutron star cores.
Three of them (EOSs N1, N2, and N3 listed in Table \ref{tab-modles})
refer to a nucleon dense matter
(neutrons, protons, and electrons), while other two
(EOSs NH1 and NH2 listed in Table \ref{tab-max})
refer to a matter containing nucleons, electrons, muons
and hyperons (of all types).

\begin{table*}[th]
\centering
\caption[]{EOSs of nucleon matter; low-mass and maximum-mass neutron
star configurations}
\label{tab-modles}
\begin{center}
  \begin{tabular}{|c|c|ccc|ccc|}
  \hline \hline
      &  &    & Low-mass  &     &   & Maximum-mass &           \\
\hline      
EOS   &  $\rho_{\rm D}$     
      &  $M$           & $\rho_{\rm c}$      & $R$
      & $M$           & $\rho_{\rm c}$      & $R$ \\
      & (g cm$^{-3}$) 
      &  ($M_\odot$)  & (g cm$^{-3}$)       & (km)
      &  ($M_\odot$)  & (g cm$^{-3}$)       & (km) \\ 
  \hline
  N1  & $7.85 \times 10^{14}$  
      & 1.1           & $6.23  \times 10^{14}$ & 13.20
      & 1.977         & $2.578 \times 10^{15}$ & 10.75 \\
  N2  & $1.298 \times 10^{15}$
      & 1.1           & $8.50  \times 10^{14} $ & 12.18
      & 1.73          & $3.25 \times 10^{15}  $ &  9.71 \\
  N3  & $1.269 \times 10^{15}$
      & 1.1           & $1.217 \times 10^{15} $ &  11.31
      & 1.46          & $4.04 \times 10^{15} $ &  8.91 \\
    \hline \hline
\end{tabular}
\end{center}
\end{table*}

\begin{table*}[th]
\centering
\caption[]{EOSs of hyperon matter; maximum-mass hyperonic
stellar configurations}
\label{tab-max}
\begin{center}
  \begin{tabular}{|c|c|c|c|ccc|}
  \hline \hline
      &      &        &     &   & Maximum-mass &           \\
\hline      
EOS   & $\rho_{\rm h}$ & $\rho_{\rm D}$      & $\rho_{\rm D1}$
      & $M$           & $\rho_{\rm c}$      & $R$ \\
      & (g cm$^{-3}$) & (g cm$^{-3}$)       & (g cm$^{-3}$)
      &  ($M_\odot$)  & (g cm$^{-3}$)       & (km) \\ 
  \hline
  NH1 & $5.495 \times 10^{14}$ 
                      & $4.06 \times 10^{14}$  & $> \rho_{\rm c}^{\rm max}$ 
      & 1.975         & $2.35   \times 10^{15}$ & 11.41 \\
  NH2 & $5.24 \times 10^{14}$           
      & $5.25 \times 10^{14}  $ &  $2.2 \times 10^{15}$ 
      & 1.740         & $3.891 \times 10^{15}  $ &   9.089  \\
    \hline \hline
\end{tabular}
\end{center}
\end{table*}

EOSs N1, N2, and N3 are
 modifications of the phenomenological EOS proposed
by \citet{pal88}. All of them allow the direct Urca process
\citep{lpph91}
to operate in a sufficiently dense 
neutron-star matter, $\rho > \rho_{\rm D}$.
The threshold density $\rho_{\rm D}$ is given in Table \ref{tab-modles}.
We present also 
the parameters of the maximum-mass configurations
(mass $M_{\rm max}$, central density $\rho_{\rm c}$, and 
radius $R$) and the parameters of low-mass neutron-star
configurations (with $M=1.1 \, \Msun$ as an example).
EOS N1 implies model I of the symmetry
energy and the compression modulus of saturated nuclear
matter $K=240$ MeV. EOSs N2 and N3 imply, respectively, $K=180$ MeV and
$K=120$ MeV, and
the symmetry energy proposed by \citet{pa92}. 
EOS N1 is rather stiff: it yields
$M_{\rm max} \approx 2 M_\odot$. EOS N2 is softer, with smaller
$M_{\rm max}$. The symmetry energy is overall smaller, which
results in a lower proton fraction and higher direct Urca threshold.
EOS N3 is an example of a quite different, soft EOS,
with $M_{\rm max} =1.46\, M_\odot$.
  
EOSs NH1 and NH2 in Table \ref{tab-max} refer to hyperon matter.
The hyperons appear at sufficiently high densities;
the density of the appearance
of the first hyperon is denoted by $\rho_\mathrm{h}$.
In the inner cores of massive hyperonic stars, a variety
of direct Urca processes are open. They 
involve nucleons and hyperons, electrons and
muons \citep{pplp92} and
give the major contribution to the neutrino luminosity
of massive stars. In Table \ref{tab-max}, $\rho_\mathrm{D}$ means
the threshold density for the first direct Urca
process (hyperonic or nucleonic).
However, if the density of hyperonic matter grows to essentially
supranuclear values, the fraction of leptons (electrons and
muons) becomes lower (they are replaced
by $\Sigma^-$ and $\Xi^-$ hyperons). Since the leptons
are important participants of Urca processes, their reduced
fraction may result in switching off these processes 
at high density. In Table \ref{tab-max}, $\rho_\mathrm{D1}$
is the highest density of operation of the last
direct Urca process.

EOS NH1 is given by the relativistic mean field
model 3 of \citet{glendenning85} and may be unrealistically
stiff for hyperonic matter ($M_{\rm max} \approx 2\,M_\odot$).
The first hyperon, $\Lambda$, appears at $\rho=\rho_\mathrm{h}$,
where nucleon direct Urca processes are already on
($\rho_\mathrm{D}<\rho_\mathrm{h}$).
The reduction of lepton fraction at high
densities is not too strong. As a result, the direct Urca
processes still
operate at the center of the maximum-mass star.

EOS NH2 is EOS2 N$\Lambda \Sigma \Xi$ of \citet{blc99}.
It is softer: $M_\mathrm{max} \approx 1.74\,M_\odot$.
The first hyperon, $\Sigma^-$, appears at the density $\rho_\mathrm{h}$
which is only slightly lower than $\rho_\mathrm{D}$.
The first direct Urca process, which opens
with increasing $\rho$, involves hyperons.
A very strong reduction of lepton fraction at high
densities switches off all direct Urca
processes at nearly the same density $\rho_\mathrm{D1}$
in central kernels of massive stars. 

\subsection{Limiting models}
\label{limiting models}

Realistic models of neutron stars
should take into account possible superfluidity
of baryons (nucleons and
hyperons) in stellar interiors (e.g., \citealp{ls01,blc99}).
Microscopic calculations give a large
scatter of critical temperatures of baryon pairing, 
depending on a
baryon-baryon interaction model and many-body theory
employed. All these calculations
indicate that superfluidity should disappear
in essentially supranuclear matter (in the centers
of massive neutron stars).
    
The theory predicts the existence of {\it three} types
of transiently accreting neutron stars
with different thermal structure (e.g., \citealt{ylh}):
{\it low-mass}, {\it medium-mass}, and {\it high-mass} stars. 
Low-mass
stars have so low central densities that the
direct Urca processes are not operative, being
either strictly forbidden ($\rho_{\rm c}<\rho_{\rm D}$)
or totally suppressed by a superfluidity
(which may be strong at not too high densities). Accordingly,
these stars have a low neutrino luminosity and 
are hottest, for a given mass accretion rate. 
The central densities of high-mass neutron stars  
are noticeably higher than $\rho_{\rm D}$
and the density, where the superfluidity dies
out and ceases to suppress the direct Urca process. 
These stars possess inner cores with fully open direct
Urca processes and greatly enhanced neutrino
luminosity. Hence, the massive stars are coldest,
for a given accretion rate. Medium-mass stars are intermediate
between the low-mass and high-mass ones.

\begin{figure}\centering
\epsfxsize=86mm
\epsffile[70 210 560 680]{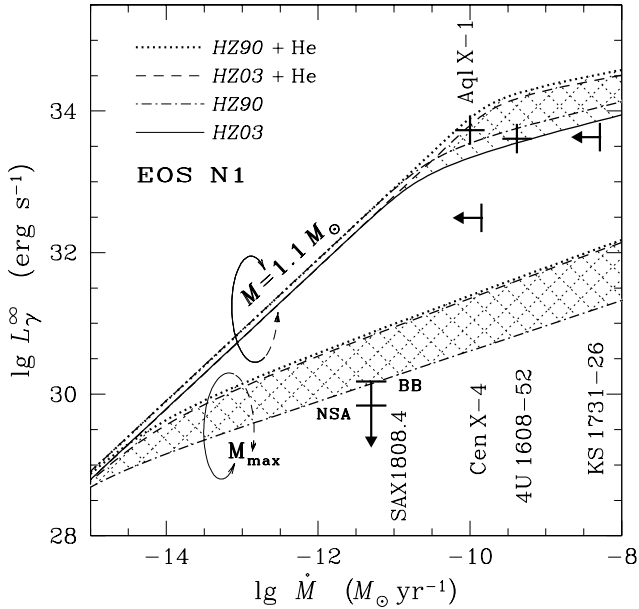}
\caption{
Theoretical quiescent thermal luminosity of neutron stars
(with EOS N1 in the cores) versus
mean mass accretion rate confronted with observations
of five SXTs. The larger upper limit
of the luminosity of \R\ assumes the blackbody (BB)
surface emission; the smaller limit -- lower
horizontal bar -- is obtained with the neutron-star
atmosphere (NSA) model. The upper curves refer to the low-mass
($1.1\,\Msun$) star (with strong proton superfluidity in the core),
while the lower curves to
the maximum-mass star. Four curves for the low-mass star
are calculated assuming HZ90 or HZ03 accreted crusts,
without or with a massive
helium surface layer; singly hatched region
shows ranges of variations of quiescent luminosity due to
different mass of helium layer for
the HZ03 crust; double hatched region is the same for
the HZ90 crust. The heating curves for the high-mass star 
are the same but the solid curve (for
the HZ03 crust without helium layer) is absent because
of the partial ionization problem (see text); it is expected to
be close to the dash-and-dot curve.
\label{fig3}}
\end{figure}

Thus, the highest heating curves (Fig.\ \ref{fig3}) correspond
to low-mass stars, and the lowest curves to high-mass stars.
Increasing the stellar mass from the lowest to the
highest values, one obtains a family of heating curves of
medium-mass stars which fill in space between
the highest and lowest curves. Any observational
point between the lowest and highest heating curves
can be explained by employed neutron-star models.   

Since the problem involves many parameters, we restrict
ourselves to the limiting cases. They
give the highest or lowest heating curves and constrain
thus theoretical values of $L_\gamma^\infty$.
The limiting low-mass and high-mass neutron-star 
configurations seem to be more robust with respect to physics
input (EOS and superfluid properties of stellar cores) 
than the medium-mass configurations.
This has been proven for cooling isolated neutron stars
(e.g., \citealt{kyg02}) and has to be true for transiently
accreting neutron stars with deep crustal heating
because of the direct correspondence of
cooling and heating problems \citep{ylh}.  
Specifically, we employ limiting values of three parameters 
listed in Table \ref{tab-limits}.

\begin{table}[!t]
\centering
\caption[]{Variation limits of three parameters of heating models}
\label{tab-limits}
\begin{center}
  \begin{tabular}{lll}
\hline \hline
\noalign{\smallskip}
   Limiting parameter & Limit 1 &  Limit 2 \\
\noalign{\smallskip}
\hline \hline
\noalign{\smallskip}
   Mass $M$       &  \parbox{7em}{$1.1\,M_\odot$\\ (low-mass star)}  & 
                 \parbox{6em}{$M_{\rm max}$ (high-mass star) } \\
\noalign{\smallskip}
\hline
\noalign{\smallskip}
   Accreted crust model  &  HZ90     &    HZ03   \\
\noalign{\smallskip}
\hline
\noalign{\smallskip}
 \parbox{10em}{Mass of He layer,\\
  $\Delta M/M =\eta / g_{14}^2 $}   &
           0           & 
           $10^{-8}$    \\
\noalign{\smallskip}
\hline \hline
\end{tabular}
\end{center}
\end{table}

The first parameter is the stellar mass; 
the limiting values refer to
low-mass and high-mass neutron stars.
For representative low-mass models,
we take the ones with $M=1.1\,M_\odot$ and nucleon cores (Table \ref{tab-modles}).
We assume further the 
strong proton superfluidity in the cores of low-mass
stars, with the critical temperature
of proton pairing equal to $5 \times 10^9$ K.
This superfluidity totally suppresses the modified Urca
process, so that the neutrino luminosity
of the star is determined by neutron-neutron bremsstrahlung
process (just as for cooling neutron stars; e.g.,
\citealt{kyg02}). Actually, our models of low-mass stars
are fairly insensitive to a specific model of the proton superfluidity
as long as the superfluidity is sufficiently strong
to suppress the modified Urca process.

Recently the neutrino emission in neutron-neutron
bremsstrahlung has been reconsidered 
by \citet{ddt03}. The authors use several realistic models
of neutron-neutron interaction and conclude that the emissivity
of the process is about four times lower than in the
simplified one-pion exchange (OPE) model. Our code incorporates
the results of \citet{fm79} obtained using the OPE model
with phenomenological correction factors. We have checked that
the emissivity provided by the code agrees with the improved
results of \citet{ddt03}, i.e., it is already
reduced by a factor of $\sim4$ with regard to the OPE model
of \citet{ddt03}.

Furthermore, we assume a weak triplet-state
neutron pairing in neutron-star cores, with maximum critical
temperature $\lesssim 2 \times 10^8$ K. This pairing has actually
no effect on thermal states of neutron stars
and can be ignored. A stronger neutron pairing would
produce a powerful neutrino emission and fast
cooling of isolated neutron stars
(e.g.,  \citealp{kyg02,cospar03}), 
in sharp disagreement with observations
of these objects. For simplicity, we neglect the
effects associated with singlet-state neutron pairing
in neutron-star crusts (e.g., \citealt{ykg01,pycg}).
To represent the high-mass stars, we use the maximum-mass
configurations (Tables \ref{tab-modles} and \ref{tab-max}). 
The effects of superfluidity are expected
to be minor in such a star,
and we neglect them in our calculations.

The second and third parameters in Table \ref{tab-limits} 
test the composition of accreted envelopes.
The second parameter specifies the composition of deep accreted crust;
we consider two limiting cases, the
HZ90 or HZ03 crusts (Sects.\ \ref{sect:intro} and \ref{sect:th-str}).
The third parameter is
the relative mass of the surface helium layer
(Sect.\ \ref{sect:th-str}). Our two limiting cases are: 
$\Delta M/M=0$ (no He layer) and
$\Delta M/M=10^{-8}$ (most massive He layer).
We will see that the effect of the helium layer on $L_\gamma^\infty$
saturates with increasing $\Delta M/M$, 
owing to which the limit of 
most massive envelope 
is actually achieved at $\Delta M/M \ll 10^{-8}$.

\subsection{Overall results (EOS N1)}
\label{results}

We start with the overall analysis,
taking EOS N1 in the stellar cores as an example.
The effects of different EOSs will be analyzed in Sects.\ \ref{aquila} 
and \ref{sax}.
Theoretical models are compared with the observations
(Sect.~\ref{observations}) in Fig.~\ref{fig3}.
Four upper heating curves refer to a low-mass
neutron star, while three lower curves refer to a high-mass
one. The solid and dashed curves are for the HZ03
crust, while the dash-and-dot and dotted curves are for the HZ90 crust.
The solid and dashed-and-dot curves are calculated assuming
no He on the surface, while the dashed and dotted
curves are for the massive He layer. We do not
present the solid curve (the HZ03 crust with the
Pd composition extended to the surface) for the maximum-mass
star. In this star, the effects of partial ionization
of Pd plasma would be very important but they are neglected
in our analysis. We expect, however, that this curve
is not too different from the dash-and-dot one 
(see below).

Our calculations
confirm the main results of the previous studies
(e.g., \citealt{ylh}): one can explain the data,
assuming that neutron star cores are composed of nucleons;
\object{Aql X-1} and 4U 1806--52 may be treated as SXTs containing low-mass
neutron stars; Cen X-4 and KS 1731--26 may be treated as
SXTs containing medium-mass neutron stars; \R\ seems to be
a source with a high-mass star; all these neutron stars
cool mainly via neutrino emission from their cores
(their photon thermal luminosity is much weaker
than the neutrino one) except possibly
\object{Aql X-1}, whose regime is intermediate
between the neutrino and photon cooling.  
In the neutrino regime, Eq.\ (\ref{thermal-balance})
reduces to $L_\mathrm{dh}^\infty(\dot{M})
\approx L_\nu^\infty(T_\mathrm{in})$, which yields
the internal temperature $T_\mathrm{in}$, while the
surface thermal emission is adjusted to this $T_\mathrm{in}$
(see \citealt{ylh} for details).

Our present results enable us to extend the consideration
of \citet{ylh}. Since we are mainly interested in the neutrino cooling
regime, the internal temperatures of neutron stars of
a given mass and accretion rate are determined by
neutrino emission and deep crustal heating model
(HZ90 or HZ03) and do not depend
on the presence of He on the surface. On the other hand,
the $T_\mathrm{eff}-T_\mathrm{in}$ relation
is actually the same for the HZ90 and HZ03 scenarios,
if massive layers of light elements are present.
A small difference between the dashed and dotted
curves is solely determined by different amount of heat
released in the HZ90 and HZ03 crusts,
see Eq.\ (\ref{Ldh}).
The heat released in the HZ90 crust
is slightly larger, so that the dotted
curves are slightly higher than their dashed counterparts.
A larger difference between the HZ90 and HZ03 crusts without
a light-element layer is mainly determined
by different thermal insulations of heat-blanketing
envelopes (Sect.\ \ref{sect:th-str}). 
The high-$Z$ (HZ03) heat blanketing envelope is
less heat transparent. Hence, the surface temperature is
smaller and the solid curve is lower than its
dot-and-dashed counterpart.
Still larger difference occurs between the scenarios
with and without helium layers. A massive helium layer
is much more heat transparent than the layer
of HZ90- or HZ03-matter. Accordingly, the dotted and dashed
heating curves go noticeably higher than the associated
dash-and-dot and solid curves.

As seen from Fig.~\ref{fig3}, the presence of light elements
on the surface of \object{Aql X-1} simplifies theoretical treatment
of \object{Aql X-1} as an SXT containing a low-mass neutron star.
The spectrum of the object is well described by hydrogen
atmosphere models, which is in line with
the assumption that the neutron star has the surface layer of
light elements. The effects of different EOSs for the interpretation of
this source are discussed in Sect.\ \ref{aquila}. 
The neutron star
in 4U 1608--52 may be treated either as a low-mass neutron
star without a massive light-element layer, or
as a medium-mass neutron star with such a layer.

The interpretation of \R\ is of special interest.
EOS N1 adopted in Fig.~\ref{fig3}
is consistent only with the black-body
thermal emission (with no massive He layer on the stellar surface)
and disagrees with
a neutron-star-atmosphere spectrum. 
Although the upper limits of $L_\gamma^\infty$ 
inferred with the black-body and neutron star atmosphere models
seem to be not very certain, our results indicate that
the neutron star in \R\ is so cold that it is barely
explained by the theory. 
In Sect.\ \ref{sax} we will show that the theoretical
explanation is relaxed if the neutron star contains
a hyperonic core. In any case, the star should have no
massive layer of light elements on the surface.

The limiting cases of no He layer
and massive He layer deserve
special comments. 
The mass of the light-element layer 
may vary from one quiescent stage to another or even
during one quiescent stage due to residual accretion.
Therefore, the heating curves for a massive He layer
and without it represent
the upper and lower limits of the quiescent thermal luminosity
of the same star \citep{bbc02}.

\begin{figure}\centering
\epsfxsize=86mm
\epsffile[50 210 570 680]{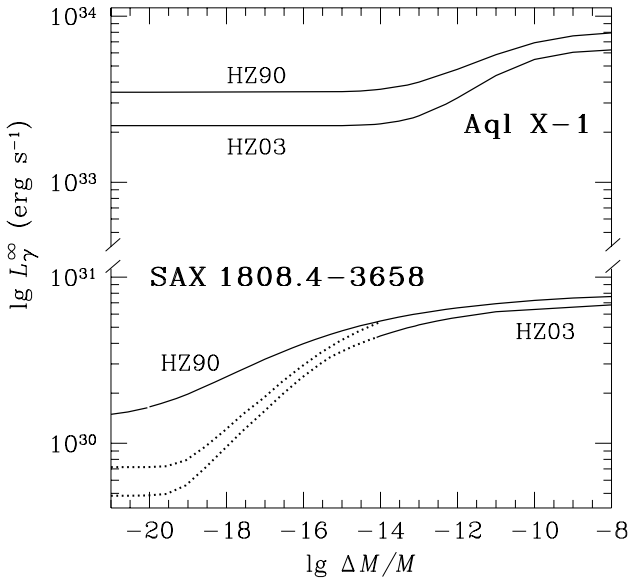}
\caption{
Theoretical quiescent thermal luminosity of neutron stars
(with EOS N1 in the cores and the HZ90 or HZ03 crusts) in \object{Aql X-1} and
\R\ (as in Fig.\ \ref{fig3}) versus
the mass of the surface He layer.
The dotted lines refer to
oversimplified (inaccurate) models which neglect partial ionization of
heavy elements (Fe or Pd) in outermost layers. 
\label{fig3a}}
\end{figure}

Figure \ref{fig3a} 
shows the surface thermal luminosity
of neutron stars with the HZ90 and HZ03 crusts
versus the mass $\Delta M$ of the helium layer. The upper curves refer to  
the low-mass star with the mean accretion rate of \object{Aql X-1}.
The lower curves correspond to the high-mass star 
with the mean accretion rate of \R.
The growth of $L_\gamma^\infty$ with increasing $\Delta M$
saturates, so that the limit of very massive He layer is
actually achieved at $\Delta M \ll 10^{-8}\, \Msun$.

Since the neutron star in \object{Aql X-1} is hot, the main
temperature gradient occurs in deep layers
of its heat-blanketing envelope. One needs at least
$\Delta M \sim 10^{-13}\,\Msun$ of helium 
to affect the thermal structure, and
$\Delta M \sim 10^{-10}\,\Msun$ of helium (extending to
$\rho \sim 10^7$ g cm$^{-3}$) to achieve the limit
of the most massive He layer. 
The increase of $\Delta M$ enhances the thermal luminosity
by a factor of 2.5--3, in agreement with the results
by \citet{bbc02}. 
It is instructive to note
that the mass of the neutron-star atmosphere 
in \object{Aql X-1} at the optical depth
$\tau=2/3$ would be $\sim 10^{-24}\, \Msun$,
while at $\tau=10$ it would be $\sim 10^{-20}\, \Msun$.
Therefore, the spectrum of thermal radiation
formed in the stellar atmosphere is affected 
by a much smaller amount of light elements than the
thermal structure of the neutron star envelope.
Note also that the plasma remains almost fully
ionized in all surface layers of this hot neutron star.

The situation with \R\ is different.
The neutron star is much colder, and the main temperature
gradient shifts to the surface. 
Variations of $L_\gamma^\infty$ with increasing $\Delta M$
reach one order of magnitude.
Even $\Delta M \sim 10^{-20}\,\Msun$ of He
(comparable with the mass of the atmosphere at $\tau \sim 10$,
less than 1 cm under the surface) 
affects the surface thermal luminosity, while the limit
of the most massive envelope is nearly achieved at
$\Delta M \sim (10^{-14}-10^{-13})\, \Msun$. At these $\Delta M$
heavy elements are partially ionized. 
As explained in Sect.\ \ref{sect:th-str}, we have taken into account
the partial ionization
of Fe but not Pd. 
Assuming artificially full ionization of Fe or Pd, we obtain the
dotted curves in Fig.\ \ref{fig3a} (the dotted curve
for the Pd/He crust at $\Delta M \lesssim 10^{-15} M$
is plotted from direct calculation rather than from
the fit expressions, Eqs.\ (\ref{fit})). This approximation
is seen to be rather inaccurate for the Fe/He envelope
at $\Delta M \lesssim 10^{-14}\,\Msun$, and is expected to be
even more inaccurate at these $\Delta M$ for the Pd/He
envelope. In this respect, cold neutron stars
(particularly, \R) can serve as laboratories for studying
ionization equilibrium of dense matter, a complicated
theoretical problem whose solution is model dependent.

\subsection{Low-mass neutron stars: \object{Aql X-1}}
\label{aquila}

The curves in Figs.~\ref{fig3} and \ref{fig3a}
are calculated for one model EOS in a
neutron star core. The calculations with
all five EOSs listed in 
Tables \ref{tab-modles} and \ref{tab-max} reveal that the
heating curves of low-mass and high-mass neutron stars are
not too sensitive to these EOSs. The heating curves of medium-mass
stars do depend on the EOS (just as for isolated 
cooling neutron stars, e.g., \citealt{kyg02})
which will be studied in
the next publication. Here, we restrict ourselves to the limiting low-mass
and high-mass models.

\begin{figure*}\centering
\epsfxsize=170mm
\epsffile[40 440 540 680]{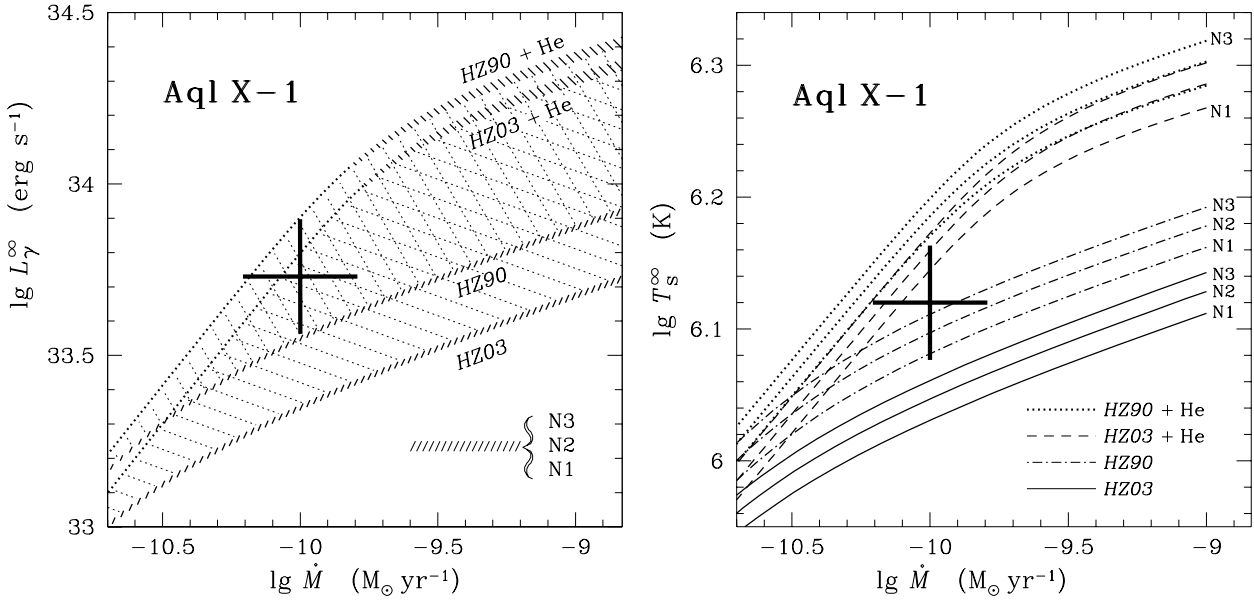}
\caption{
Quiescent thermal luminosity (left panel) or
effective surface temperature (right panel)
vs.\ $\dot{M}$ for a $1.1\, \Msun$ star with a nucleon core
(containing strongly superfluid protons) confronted with
observations of \object{Aql X-1}. 
Set of four heating curves:
with HZ90 or HZ03 crusts, with
and without massive helium layer, is shown for 
the three EOSs in the stellar cores (EOSs N1, N2 and N3).
Fixing the accreted crust model,
we obtain almost the same luminosities (the curves
for EOSs N1, N2, and N3 almost coincide; left panel) but
different effective temperatures (right panel).
Hatched regions in the left panel display ranges of
possible variations of $L_\gamma^\infty$ due to
variable thickness of the He layer for the HZ90 and HZ03 crusts.
\label{fig4}}
\end{figure*}

Let us start with the low-mass stars. 
Figure~\ref{fig4} shows the heating curves of $1.1\,M_\odot$
neutron stars with the cores described by EOSs N1, N2, and
N3
and the envelopes composed of either the HZ90 or
HZ03 matter, without and with He layers.
The left panel displays our traditional
heating curves, $L_\gamma^\infty(\dot{M})$,
while the right panel gives $T_\mathrm{eff}^\infty(\dot{M})$
for the same scenarios.

The left and right panels of Fig.~\ref{fig4} are seen to be
different. If we fix a model of accreted
crust and the value of $\Delta M/M$, the $L_\gamma^\infty(\dot{M})$
heating curves are almost \textit{universal} -- insensitive to
the EOS in the stellar core. We plot them by wide
hatched lines in the left panel: the curves for different
EOSs would be almost indistinguishable. 
In addition, we have calculated the 
heating curves for a $M=1.3 \, \Msun$ neutron star with EOS N1
(not presented in Fig.\ \ref{fig4})
and check that they are also indistinguishable from
corresponding $1.1 \, \Msun$ curves.
On the other hand,
the $T_\mathrm{eff}^\infty(\dot{M})$ heating curves
noticeably depend on the EOS: the
luminosity $L_\gamma^\infty(\dot{M})$ is the same, 
but stellar radii are different (Table \ref{tab-modles}), hence
different $T_\mathrm{eff}^\infty(\dot{M})$.
Notice that the theory of
cooling isolated neutron stars predicts just the
opposite property of the cooling curves ($L_\gamma^\infty$
or $T_\mathrm{eff}^\infty$ versus stellar age $t$):
the cooling curves $T_\mathrm{eff}^\infty(t)$
of low-mass neutron stars are almost universal 
(e.g., \citealt{kyg02}), whereas the curves
$L_\gamma^\infty(t)$ should noticeably depend on the EOS.

The upper heating curves in Fig.~\ref{fig4} correspond to the stars with
massive light-element layers, while the lower
curves are calculated assuming no such layers.
The difference between the lower and upper curves is
quite pronounced.
As seen from Fig.~\ref{fig4}, 
observations of \object{Aql X-1} with simultaneous stringent determinations
of $L_\gamma^\infty$ and $T_\mathrm{eff}^\infty$ have potential
to constrain the EOS in the core of low-mass neutron stars and
to determine the composition of the  surface layers.
On the other hand, the neutron star is so hot
that it is nearly at the edge of theoretical ability to
explain hot stars. Were the quiescent
thermal luminosity $L_\gamma^\infty 
\gtrsim 10^{34}$ erg s$^{-1}$ 
detected in future observations
of \object{Aql X-1}, it could not be explained by deep crustal heating.
This may be regarded as an additional test of the deep crustal heating
mechanism.  

\subsection{The coldest massive neutron star: \R}
\label{sax}

\begin{figure}\centering
\epsfxsize=86mm
\epsffile[80 210 550 680]{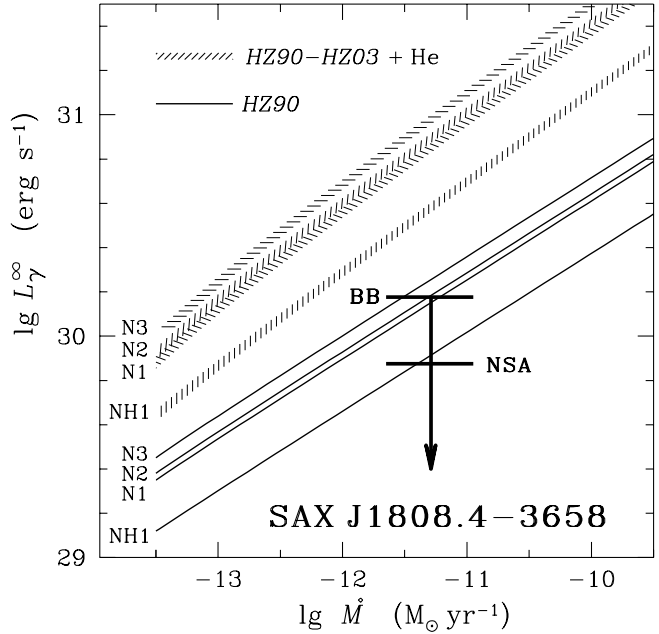}
\caption{
Quiescent thermal luminosity
vs.\ $\dot{M}$ for maximum-mass neutron-star models
confronted with
observations of SAX J1808.4--3658. Based on 15 curves for
five EOSs in the stellar core in the presence of
massive helium surface layers (HZ90 and HZ03 crusts, hatched lines)
and in the absence of these layers (HZ90 crust, solid lines). 
          The EOS types are marked to the left of the lines.
The curves for EOSs N2 and NH2
are almost indistinguishable.
Taking the massive helium layer and
a fixed EOS, we have two almost coincident curves
for HZ90 and HZ03 crusts,
plotted as a wide hatched line. 
\label{fig5}}
\end{figure}

Let us turn to the high-mass neutron-star models.
Figure~\ref{fig5} displays the heating curves of maximum-mass neutron stars
with five EOSs in their cores, the HZ90 or
HZ03 composition of the crust, with or without
helium surface layers.
If we fix the composition of the crust and the helium mass,
then the difference between the thermal luminosities of neutron stars
with four EOSs in the core (nucleon EOSs N1, N2, and N3,
and hyperon EOS NH2) is small. Thus, just as for
low-mass neutron stars (Sect.\ \ref{aquila}),
the heating curves $L_\gamma^\infty(\dot{M})$ are
{\it nearly universal} (while the curves $T_\mathrm{eff}^\infty(\dot{M})$
would be much more different). 

Employing the four EOSs (N1, N2, N3, and NH2), we come to
the same conclusions as in Sect.\ \ref{results}:
the theory is consistent with the upper limit of $L_\gamma^\infty$
inferred using the blackbody model 
(without any light-element layer on the surface), and it is inconsistent
with the limit of $L_\gamma^\infty$ inferred with the
neutron-star atmosphere model. The appearance of a helium layer
can raise the quiescent thermal luminosity up to an order
of magnitude.

Nevertheless, the stars with the hyperon EOS NH1
{\it disobey the universality}: they are 
noticeably colder. Their coldness is
explained by very strong neutrino emission from
from the hyperonic stellar core.
As discussed in Sect.\ \ref{EOS},
the fraction of leptons in the central region of the
maximum-mass neutron star with EOS NH1 is sufficient to keep
all direct Urca processes open everywhere within the inner
core, including the stellar center. The neutrino emission
becomes exceptionally intense, and the star very
cold. 
Such a cold high-mass hyperonic star without a massive
helium surface layer is in much better agreement
with the observations of \R. As seen from Fig.\ \ref{fig5},
in this case the theory is not only consistent
with the upper blackbody limit on $L_\gamma^\infty$, 
but becomes in reasonable
agreement with the upper limit inferred using the
atmosphere model (which is not the case for other EOSs).
This opens a potential possibility to discriminate
between the different nucleon/hyperon EOSs
and 
gives a tentative indication that the neutron star
in \R\ contains a hyperonic core with a not too low
lepton fraction. 

As discussed in Sect.\ \ref{EOS},
the lepton fraction for EOS NH2
decreases very rapidly with increasing density. This switches
off all direct Urca processes at $\rho> \rho_\mathrm{D1}$,
in the central kernel of the maximum-mass
neutron star, reducing thus the neutrino emission
to the level of neutron stars with nucleon cores.

Thus, further observations of \R\ in quiescent 
are highly desirable. Future detections or constraints
of the quiescent thermal luminosity $L_\gamma^\infty$
would be most important, especially those which give lower
values of $L_\gamma^\infty$ than the
present ones. For instance the value
of $L_\gamma^\infty \approx 6 \times 10^{29}$ erg s$^{-1}$
would definitely imply a massive hyperon neutron star with 
not too small fraction of leptons in its
center (to keep the direct Urca processes open) and
without any massive surface layer
of light elements.
Lower values of $L_\gamma^\infty$
could not be explained by the current model.

\section{Conclusions}
\label{conclusions}

We have considered (Sect.\ \ref{sect:th-str}) the growth of temperature
within the heat-blanketing envelope of a
transiently accreting neutron star
in a quiescent state. We have analyzed two basic models 
HZ90 and HZ03 of 
the accreted crust, calculated by
\citet{HZ90,HZ03}. 
In all cases we consider the
possible presence of 
a thin ($\Delta M \lesssim 10^{-8}\, \Msun$) layer of
light elements (H or He) on the surface. 
We have calculated
the relations between the internal and surface
temperatures of neutron stars with the HZ90 or HZ03 crusts
and fitted the results by simple expressions.

Using these results, we have modeled (Sect.\ \ref{sect:th-states}) 
thermal
states of transiently accreting neutron stars in SXTs,
assuming that these states are
regulated by deep crustal heating in accreted matter.
We have considered five model EOSs of nucleon
or nucleon-hyperon matter in neutron star cores,
representative models of low-mass and high-mass neutron
stars, HZ90 and HZ03 models of stellar crusts,
without light elements on the stellar surfaces or
with maximum amount of light elements. 
The results give the upper and lower limits of
the quiescent thermal luminosity
of SXTs, depending
on the amount of light elements at the neutron-star surface
in a particular quiescent period.

We have compared
the theory with observations of five SXTs. The most
important are two sources, \object{Aql X-1} and SAX J1808.4--3658.
\object{Aql X-1} can be treated as a low-mass, warm neutron star.
Its future observations may constrain
the EOS in the nucleon core of a low-mass neutron star, 
elucidate the composition of
accreted matter, and test the deep crustal heating hypothesis.
The second source, SAX J1808.4--3658, can be treated
as a very cold massive neutron star with nucleon
or nucleon-hyperon core;
a hyperonic core with not too low
fraction of electrons is more preferable. 
Future observations of \R\ in
quiescence may enable one to distinguish between
the EOSs in massive nucleon/hyperon stellar cores
and check the ionization models of heavy-element plasma
in surface layers of neutron stars. 

The assumption that the quiescent thermal emission of SXTs is
produced by the deep crustal heating 
\citep{bbr98} remains still a hypothesis.
However, the theory of deep crustal heating \citep{HZ90,HZ03}
is solid: accreting neutron stars
should be heated from inside and this effect
cannot be avoided.

\begin{acknowledgements}
We are grateful to R.\ V.\ E. Lovelace for careful reading the
manuscript and useful remarks.
This work was supported in part by the
RFBR (grants No.\ 02-02-17668 and 03-07-90200).
\end{acknowledgements}

\end{document}